\documentclass[prd,twocolumn,floatfix,preprintnumbers,letterpaper]{revtex4}
\usepackage{graphicx}
\usepackage{amsmath,amssymb}
\input{epsf}
\usepackage{epsf}

\newcommand {\ga} {\ {\raise-.5ex\hbox{$\buildrel>\over\sim$}}\ }
\newcommand {\la} {\ {\raise-.5ex\hbox{$\buildrel<\over\sim$}}\ } 

\begin{document}

\title{Thawing quintessence with a nearly flat potential}
\author{Robert J. Scherrer}
\affiliation{Department of Physics and Astronomy, Vanderbilt University,
Nashville, TN  ~~37235}
\author{A.A. Sen}
\affiliation{Center For Theoretical Physics,
Jamia Millia Islamia, New Delhi 110025, India}

\begin{abstract}
The thawing quintessence model with a nearly flat
potential provides a natural
mechanism to produce an equation of state parameter, $w$, close
to $-1$ today.
We examine the behavior of such models
for the case in which
the potential satisfies the slow roll conditions:  $[(1/V)(dV/d\phi)]^2 \ll 1$
and $(1/V)(d^2 V/d\phi^2) \ll 1$, and we derive the analog
of the slow-roll approximation for the case in which both
matter and a scalar field contribute to the density.
We show that in this limit, all such models converge to
a unique relation between $1+w$, $\Omega_\phi$, and the
initial value of
$(1/V)(dV/d\phi)$.  We derive this relation, and use it to
determine the corresponding expression for $w(a)$, which depends
only on the present-day values for $w$ and $\Omega_\phi$.
For a variety of potentials, our limiting expression
for $w(a)$ is typically accurate to within $\delta w \la 0.005$ for $w<-0.9$.
For redshift $z \la 1$, $w(a)$ is well-fit by the Chevallier-Polarski-Linder
parametrization, in which $w(a)$ is a linear function of $a$.

\end{abstract}

\maketitle

\section{Introduction}

Observational evidence \cite{Knop,Riess1}
indicates that roughly 70\% of the energy density in the
universe is in the form of an exotic, negative-pressure component,
dubbed dark energy.  (See Ref. \cite{Copeland}
for a recent review).  The observational bounds on the properties
of the dark energy have continued to tighten.  Taking
$w$ to be the ratio of
pressure to density for the dark energy:
\begin{equation}
\label{w}
w = p_{DE}/\rho_{DE},
\end{equation}
recent observational constraints are typically
$-1.1 \la w \la -0.9$ when $w$ is assumed constant
(see, e.g., \cite{Wood-Vasey,Davis}
and references therein).

We can consider two possibilities.  If the measured value
of $w$ continues to converge to a value arbitrarily close to $-1$,
then it is most reasonable to assume a cosmological constant
(a conclusion supported by both the Akaike information criterion \cite{Liddle}
and common sense).  On the other hand, it is conceivable that
the observations will converge on a value of $w$ very close to, but
not exactly equal to, $-1$.  In this case, we must consider how such
a dark energy equation of state might arise.

One possibility,
dubbed quintessence, is a model in which the dark energy
arises from a scalar
field \cite{ratra,turner,caldwelletal,liddle,zlatev}.  Caldwell
and Linder \cite{CL}
showed that quintessence models in which the
scalar field potential asymptotically approaches zero
can be divided naturally into two categories, which they dubbed
``freezing" and ``thawing" models, with quite different
behaviors.  Thawing models have a value
of $w$ which begins near $-1$ and increases with time, while
freezing models have a value of $w$ which decreases with time, with
an asymptotic value that depends on the shape of the potential.
(If the observations converge to a value of $w$ less than $-1$, more
exotic models must be considered.  We will not consider
this possibility here).

Thawing models with a nearly flat
potential provide a natural way to produce
a value of $w$ that is close to, but not exactly equal to $-1$,
since the field begins with $w \approx -1$, and $w$ increases
only slightly up to the present.  Furthermore, with a nearly flat
potential and
$w \approx -1$, the dynamics of quintessence
are considerably simplified.  In this paper, we show
that all such models converge to a single unique evolution.
In addition to providing a plausible
model for $w$ near $-1$, such models (since they all converge
to a single type of behavior) can
serve as a useful set of fiducial models
that can be compared to $\Lambda$CDM.

In the next section, we reexamine thawing quintessence models and
outline the arguments for
these models.
In Sec. 3, we present a general description of the
behavior of such models.  Our main results
are given in equations (\ref{mainresult}) and (\ref{wpred}).
In Sec. 4, we discuss the
arguments against thawing quintessence models.
Our conclusions are summarized briefly in Sec. 5.

\section{The case for thawing quintessence}

We will assume that the dark energy is provided by a minimally-coupled
scalar field, $\phi$, with equation of motion given by
\begin{equation}
\label{motionq}
\ddot{\phi}+ 3H\dot{\phi} + \frac{dV}{d\phi} =0,
\end{equation}
where the Hubble parameter $H$ is given by
\begin{equation}
\label{H}
H = \left(\frac{\dot{a}}{a}\right) = \sqrt{\rho/3}.
\end{equation}
Here $a$ is the scale factor, $\rho$ is the total density, and we work in units
for which $8 \pi G = 1$.
Equation (\ref{motionq}) indicates
that the field rolls downhill in the potential $V(\phi)$,
but its motion is damped by a term proportional to $H$.

The pressure and density of the
scalar field are given by
\begin{equation}
p = \frac{\dot \phi^2}{2} - V(\phi),
\end{equation}
and
\begin{equation}
\label{rhodense}
\rho = \frac{\dot \phi^2}{2} + V(\phi),
\end{equation}
respectively, and the equation of state parameter, $w$,
is given by equation (\ref{w}).

As noted above, observations suggest a value of $w$ near $-1$.  Caldwell
and Linder 
\cite{CL} noted there are basically two ways to achieve such a result
(see also Ref. \cite{Linder} for a more detailed discussion of some
of these issues).  In the first case, freezing potentials, the field is initially
rolling down the potential with $w \ne -1$, but it slows with time,
driving $w$ toward $-1$.  (As emphasized in Ref. \cite{Linder}, the tracking
models introduced in Ref. \cite{zlatev} are a subset of freezing models,
but not all freezing models display tracking behavior).
In thawing models, 
the field is initially nearly frozen at some value $\phi_0$, with $w=-1$.
Then as $H$ decreases, the field rolls down
the potential, and
$w$ increases with time.  

It is possible to produce $w$ near $-1$ at the present
using either freezing or thawing models.  Chongchitnan and Efstathiou \cite{Chong}
used Monte Carlo simulations to derive a set of
potentials that yield $w$ very close to $-1$ today.  They found
two classes of acceptable solutions:  very flat potentials, and models
in which the field evolves from a region with very steep slope in the
potential to a region in which the potential
is roughly flat.
While neither type of model can be ruled out, we feel that the models
with a nearly flat potential are clearly a more natural way to produce
the desired present-day value of $w$ near $-1$.

A similar naturalness issue was raised by Bludman
\cite{Bludman}, who argued that the only way to achieve
tracking models with $w$ near $-1$ today
was for the
model to contain a sharp change in the curvature of the potential
at the present.  Thus, we are presented with a double coincidence
problem:  why should the field be entering this special region of
the potential at the same time that the dark energy
is coming to dominate the matter, and why are both of these happening
right now?  In thawing models with a nearly flat potential, on
the other hand, $w$ never deviates
very far from $-1$.

Another argument in favor of these thawing models is that we already
have strong evidence that the universe at one time underwent a period
of vacuum energy domination (inflation).  Many models for inflation
correspond to the sort of thawing models examined here \cite{LL};
the scalar field initially has $w = -1$, but then rolls
downhill to terminate inflation.

Finally, Griest \cite{Griest} has suggested a solution to the
coincidence problem involving thawing fields.  In his model,
the universe contains a variety of scalar fields with
various initial energy scales $V_i(\phi_0), i = 1,2,3,...$.  As the matter or
radiation density drops below a given $V_i(\phi_0)$, the universe
undergoes a period of dark energy domination, but the field
then thaws and slides down the potential, allowing matter or radiation
to dominate again.  Given enough of these fields, it would not
be suprising to find ourselves in an epoch in which one of them
is just beginning to dominate at present (see also the somewhat
different model of Ref. \cite{Kaplinghat}).

None of these arguments proves, of course, that a universe with
$w$ close to $-1$ (but not equal to $-1$) must involve a thawing
quintessence field with a nearly flat potential.  However, they do indicate that such models
are worthy of further study.

\section{Evolution of Thawing Quintessence with Nearly Flat Potential}

We will assume a scalar field with initial
value $\phi_0$ in a nearly flat potential $V(\phi)$.  Specifically,
we will assume that
at $\phi = \phi_0$,
the field satisfies the slow-roll conditions:
\begin{equation}
\label{slow1}
\left(\frac{1}{V} \frac{dV}{d\phi}\right)^2 \ll 1,
\end{equation}
and
\begin{equation}
\label{slow2}
\frac{1}{V}\frac{d^2 V}{d\phi^2} \ll 1.
\end{equation}
The latter condition corresponds to a mass scale
$m_\phi \ll \sqrt{V(\phi_0)}$.  Setting $V(\phi_0)$ roughly
equal to the dark energy density at the present, we get
$m_\phi \ll 10^{-33}$ eV.  This is the same (unnaturally small)
mass that occurs generically in quintessence models for dark energy.

In analyzing models for inflation, it is usually assumed that
the scalar field dominates the expansion,
and that $V(\phi) \gg \dot \phi^2/2$.
With these assumptions, along with the flatness of the potential
given by equations (\ref{slow1}) and (\ref{slow2}),
it can be shown that the $\ddot \phi$ term in equation
(\ref{motionq}) can be neglected, yielding
the simple equation $3 H \dot \phi = -dV/d\phi$ (see,
e.g., Ref. \cite{LL}).  This is called the slow-roll approximation.

It is well-known that the slow-roll approximation fails for the
case of quintessence \cite{Linder,Bludman,Capone}.  The basic
reason is that the slow-roll approximation requires the scalar
field to dominate the expansion.  However, this is never the
case for quintessence, since matter always contributes significantly
to the total density.  However, nothing prevents us from
assuming the slow roll conditions on the potential (equations
\ref{slow1} and \ref{slow2}) along with the requirement that
$w$ be close to $-1$ today.  Effectively,
we are deriving the analog of the slow-roll approximation for
the case where the expansion is not dominated by the scalar field.

At the late times which are of interest here, the universe is dominated
by dark energy (assumed to arise from a scalar field)
and nonrelativistic matter; we can neglect the radiation
component.  We assume a flat universe
containing only matter and a scalar field, so that
$\Omega_\phi + \Omega_M = 1$. Then equations (\ref{motionq})
and (\ref{H}) can be rewritten in
terms of the variables $x$, $y$, and $\lambda$, defined by
\begin{eqnarray}
\label{xevol}
x &=& \phi^\prime/\sqrt{6}, \\
y &=& \sqrt{V(\phi)/3H^2}, \\
\lambda &=& -\frac{1}{V}\frac{dV}{d\phi},
\end{eqnarray}
and the prime will always denote the derivative with respect to $\ln a$:
e.g., $\phi^\prime \equiv a(d\phi/da)$.
(This discussion, from equation \ref{xevol} through equation \ref{Gamma},
is taken from Refs. \cite{CLW,MP,Ng}).

Then $x^2$ gives the contribution of the kinetic energy of the scalar
field to $\Omega_\phi$, and $y^2$ gives the contribution of the
potential energy, so that
\begin{equation}
\label{Om}
\Omega_\phi = x^2 + y^2,
\end{equation}
while the equation of state is
\begin{equation}
\label{gamma}
\gamma \equiv 1+w = \frac{2x^2}{x^2 + y^2}.
\end{equation}
It is convenient to work in terms of $\gamma$, since we are
interested in models for which $w$ is near $-1$,
so $\gamma$ is near zero,
and we can then expand quantities of interest to lowest order in $\gamma$.
Equations (\ref{motionq}) and (\ref{H}),
in a universe containing only matter and a scalar field,
become
\begin{eqnarray}
x^\prime &=& -3x + \lambda\sqrt{\frac{3}{2}}y^2 + \frac{3}{2}x[1 + x^2-y^2],\\
y^\prime &=& -\lambda\sqrt{\frac{3}{2}}xy + \frac{3}{2}y[1+x^2-y^2],\\
\lambda^\prime &=& - \sqrt{6} \lambda^2(\Gamma - 1) x,
\end{eqnarray}
where
\begin{equation}
\label{Gamma}
\Gamma \equiv V \frac{d^2 V}{d\phi^2}/\left(\frac{dV}{d\phi} \right)^2.
\end{equation}
We now rewrite these equations, changing the dependent variables from
$x$ and $y$ to the observable quantities $\Omega_\phi$
and $\gamma$ given by equations (\ref{Om}) and (\ref{gamma}).
To make this transformation, we assume that $x^\prime > 0$; our
results generalize trivially to the opposite case.
We obtain:
\begin{eqnarray}
\label{gammaprime}
\gamma^\prime &=& -3\gamma(2-\gamma) + \lambda(2-\gamma)\sqrt{3 \gamma
\Omega_\phi},\\
\label{Omegaprime}
\Omega_\phi^\prime &=& 3(1-\gamma)\Omega_\phi(1-\Omega_\phi),\\
\label{lambda}
\lambda^\prime &=& - \sqrt{3}\lambda^2(\Gamma-1)\sqrt{\gamma \Omega_\phi}.
\end{eqnarray}
Note that equation (\ref{gammaprime}) also follows, in a trivial
way, from the expression for $w^\prime$ given in Ref. \cite{Linder}.
Finally, we will see that the equations simplify if we transform our dependent variable
from $a$ to $\Omega_\phi(a)$.  This gives us
\begin{equation}
\label{exact}
\frac{d\gamma}{d\Omega_\phi} = \frac{\gamma^\prime}{\Omega_\phi^\prime}
= \frac{-3\gamma(2-\gamma) + \lambda(2-\gamma)\sqrt{3 \gamma \Omega_\phi}}
{3(1 - \gamma)\Omega_\phi(1-\Omega_\phi)}.
\end{equation}
This change of variables is valid only if $\Omega_\phi$
is a monotonic function of the scale factor; it breaks
down at any point where $d\Omega_\phi/da = 0$.  This condition
is satisfied for most quintessence models and for
all of the models we consider here; it is not satisfied,
for example, in models in which $\Omega_\phi$ oscillates
in time \cite{Kaplinghat}.

Equations (\ref{lambda}) and (\ref{exact}) are an exact description of the scalar
field evolution for $x^\prime > 0$, but they do not yield any simple solution.
At this point, we make two assumptions.  Our first assumption is
that $\gamma \ll 1$, corresponding
to $w$ near $-1$.  The second assumption is that $\lambda$ is approximately
constant, so that
\begin{equation}
\label{lamb0}
\lambda = \lambda_0 = -(1/V)(dV/d\phi)\biggr|_{\phi =\phi_0},
\end{equation}
i.e., $\lambda_0$ is the value of $\lambda$ at the initial value
of the scalar field $\phi_0$ before it begins to roll down the potential.
Equation (\ref{lamb0}) follows from the slow-roll
conditions, equations (\ref{slow1}) and (\ref{slow2}), as we will show
at the end of this calculation.
Replacing $\lambda$ with $\lambda_0$ and retaining terms to lowest order in $\gamma$
in equation (\ref{exact}) yields the following:
\begin{equation}
\frac{d\gamma}{d\Omega_\phi} = - \frac{2 \gamma}{\Omega_\phi (1-\Omega_\phi)}
+ \frac{2}{3} \lambda_0 \frac{\sqrt{3 \gamma}}{(1-\Omega_\phi)\sqrt{\Omega_\phi}}.
\end{equation}
This equation can be transformed into a linear differential
equation with the change of variables $s^2 = \gamma$, and the resulting
equation can be solved exactly.  For the models of interest here,
we have the boundary condition $\gamma = 0$ at
$\Omega_\phi = 0$.  The resulting solution (reexpressed in terms of $w$)
is
\begin{eqnarray}
1 &+& w = \frac{\lambda_0^2}{3}\left[\frac{1}{\sqrt{\Omega_\phi}}
- \left(\frac{1}{\Omega_\phi} - 1 \right) \tanh^{-1}\sqrt{\Omega_\phi}\right]^2,
\nonumber\\
\label{mainresult}
&=& \frac{\lambda_0^2}{3}\left[\frac{1}{\sqrt{\Omega_\phi}}
- \frac{1}{2}\left(\frac{1}{\Omega_\phi} - 1 \right)
\ln \left(\frac{1+\sqrt{\Omega_\phi}}
{1-\sqrt{\Omega_\phi}} \right)\right]^2.
\end{eqnarray}
Equation (\ref{mainresult}), along with the
corresponding result for $w(a)$
derived below, is our main result.  It shows that for sufficiently
flat potentials, all thawing quintessence models with $w$ near
$-1$ approach a single generic behavior, with $w(a)$ determined entirely
by $\Omega_\phi(a)$ and the (constant) initial value of $(1/V)(dV/d\phi)$.
A graph of this generic relationship between $w$, $\Omega_\phi$,
and $\lambda_0$ is given in Fig. 1.

Equation (\ref{mainresult}) shows that $1+w \sim O(\lambda_0^2)$.
Thus, our first slow-roll condition (equation \ref{slow1}) insures
that $1+w \ll 1$, as desired.  The condition that $\lambda$
be nearly constant up to the present
day can be quantified by requiring $|\lambda^\prime/\lambda| \ll 1$. Taking $\gamma$ to be of order
$\lambda^2$ in equation (\ref{lambda}), 
we obtain the condition
\begin{equation}
\frac{1}{V}\frac{d^2 V}{d\phi^2} -
\left(\frac{1}{V} \frac{dV}{d\phi}\right)^2 \ll 1.
\end{equation}
The two slow-roll conditions, taken together, insure that this condition is satisfied.
\begin{figure}[t]
\centerline{\epsfxsize=3.7truein\epsfbox{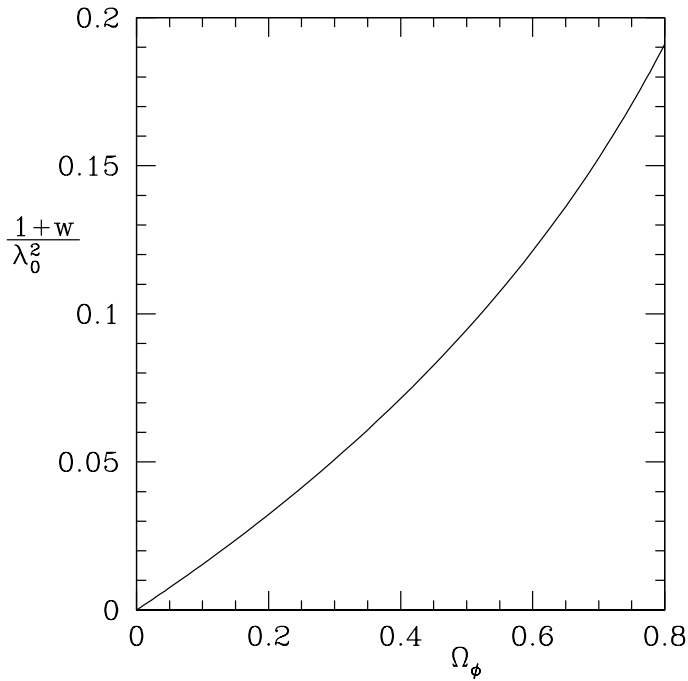}}
\caption{The value of $(1+w)/\lambda_0^2$ as a function of $\Omega_\phi$
in thawing quintessence models with a nearly flat potential and $w$ near $-1$.  Here
$w$ and $\Omega_\phi$ are functions of the scale factor $a$, while
$\lambda_0$ is the (constant) initial value of $-(1/V)(dV/d\phi)$.}
\end{figure}

A sufficiently accurate determination of
the present-day values of $w$ and $\Omega_\phi$ ($w_0$ and
$\Omega_{\phi 0}$, respectively)
uniquely determines
the value of
$\lambda_0$ for these models.
For example, for $\Omega_{\phi 0} = 0.7$ and $w_0 = -0.9$,
we obtain $\lambda_0 = 0.8$.

The way in which arbitrary potentials satisfying the slow-roll conditions
converge to equation (\ref{mainresult}) is illustrated in Fig. 2.
\begin{figure}[t]
\centerline{\epsfxsize=3.7truein\epsfbox{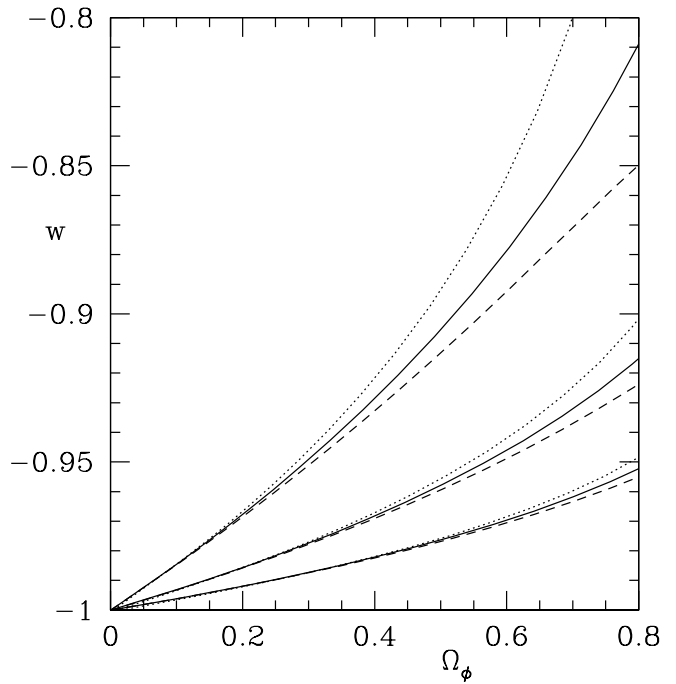}}
\caption{A comparison between $w(\Omega_\phi)$ for $V(\phi) = \phi^2$ (dotted
curve), $V(\phi) = \phi^{-2}$ (dashed curve) and our analytic result
for $w(\Omega_\phi)$ (solid curve) for (top to bottom)
$\lambda_0 = 1$, $\lambda_0 = 2/3$, and $\lambda_0 = 1/2$.}
\end{figure}
In this figure, the solid curve gives the behavior for $w(\Omega_\phi)$
predicted by equation (\ref{mainresult}), while the dotted and
dashed curves give the true evolution for the potentials
$V = \phi^2$ and $V = \phi^{-2}$, respectively, where we choose
the initial value of $\phi$ such that $\lambda_0 = 1$, $2/3$, $1/2$.
As expected, agreement is poor for $\lambda_0 = 1$ and improves
for smaller values of $\lambda_0$.  Since we dropped terms
of order $1+w$ in deriving equation (\ref{mainresult}),
we expect the fractional error in $1+w$ to be on the order of $1+w$.
For $w < -0.9$, this translates into an error in $w$ of
$\delta w \la 0.01$, which is apparent in Fig. 2.

The behavior of the $\phi^{-2}$ potential demonstrates an important
point:  while negative power law potentials usually give rise to ``freezing"
models \cite{ratra,liddle,zlatev}, they can be made to act as
thawing models by an appropriate choice of $\phi_0$.  For example,
when $V(\phi) = \phi^{-n}$,
if $\phi_0 \gg n$, then equations (\ref{slow1}) and (\ref{slow2})
are satisfied, and the model behaves like a thawing model.
This shows that {\it any} potential can give rise to the type of models
discussed here, as long as $V(\phi)$ has a region over which equations
(\ref{slow1}) and (\ref{slow2}) apply.

\begin{figure}[t]
\centerline{\epsfxsize=3.7truein\epsfbox{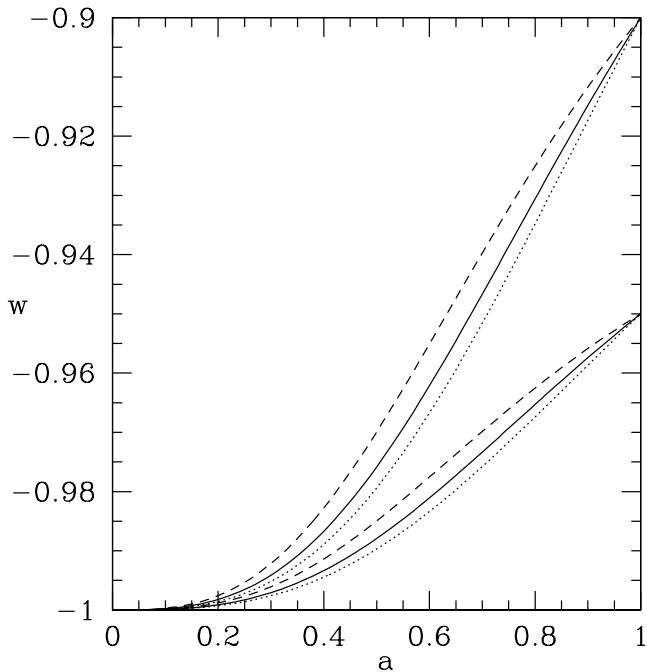}}
\caption{Our analytic result for the evolution of $w$ as a function of the
scale factor,
$a$, normalized to $a=1$ at the present,
in thawing quintessence models with a nearly flat potential and $w$ near $-1$,
for $\Omega_{\phi 0} = 0.8$ (dashed), $\Omega_{\phi 0} = 0.7$ (solid) and
$\Omega_{\phi 0} = 0.6$ (dotted).  Upper three curves
are for $w_0 = -0.9$; lower three curves are for $w_0 = -0.95$.}
\end{figure}

We can use equation (\ref{Omegaprime}) to solve for $\Omega_\phi$
as a function of $a$ and thus determine $w(a)$.  Taking
the limit $\gamma \ll 1$ in equation (\ref{Omegaprime}) gives
$\Omega_\phi^\prime = 3\Omega_\phi(1-\Omega_\phi)$,
with solution
\begin{equation}
\label{Oma}
\Omega_\phi = \left[1 + \left(\Omega_{\phi 0}^{-1} - 1 \right)a^{-3}
\right]^{-1},
\end{equation}
where $\Omega_{\phi 0}$ is the present-day
value of $\Omega_\phi$, and we take $a=1$ at the present.
Equation (\ref{Oma}) is identical to the expression for $\Omega_\Lambda$ as a function of $a$
in the $\Lambda$CDM model, which is not surprising, as we
are taking $w$ near $-1$ (see also Ref. \cite{Crit}).
Equations (\ref{mainresult}) and (\ref{Oma}) together give an
explicit expression for $w$ as a function of $a$.  Assuming a particular
value $w_0$ for the present-day value of $w$,
we can then eliminate $\lambda_0$ from this
expression, so that $w(a)$ is a unique function of $w_0$
and $\Omega_{\phi 0}$.  We obtain:
\begin{eqnarray}
1 + w = (1+ w_0)\Biggl[ \sqrt{1 + (\Omega_{\phi 0}^{-1} - 1)a^{-3}}\nonumber\\
- (\Omega_{\phi 0}^{-1} - 1)a^{-3} \tanh^{-1} \frac{1}{\sqrt{1 + (\Omega_{\phi 0}^{-1}-1)a^{-3}}}
\Biggr]^2\nonumber\\
\label{wpred}
\times \left[\frac{1}{\sqrt{\Omega_{\phi 0}}}
- \left(\frac{1}{\Omega_{\phi 0}} - 1 \right)
\tanh^{-1}\sqrt{\Omega_{\phi 0}}\right]^{-2}.
\end{eqnarray}
A graph of $w(a)$ is shown in Fig. 3 for several values of $w_0$ and
$\Omega_{\phi 0}$.  Note that $w(a)$ depends primarily on $w_0$ and is not very sensitive
to the value of $\Omega_{\phi 0}$ for $0.6 < \Omega_{\phi 0} < 0.8$.
Also, $w(a)$ is a nearly linear function of $a$
for $a$ between $0.5$ and $1$ (redshift $z \la 1$), although
the linear behavior breaks down for $z > 1$.  The $z \la 1$ behavior agrees
well with the Chevallier-Polarski-Linder
parametrization \cite{CP,Lindp}, in which $w(a)$ is taken to have
the form $w(a) = w_0 + w_a(1-a)$, although equation (\ref{wpred}) does
not provide any particular insight into the origin of this linear behavior.
However, in our case $w_a$ is not a free parameter, but depends on
$w_0$ and $\Omega_{\phi 0}$; for a fixed value of $\Omega_{\phi 0}$, equation (\ref{wpred})
corresponds to a one-parameter family of models.  For instance,
for $\Omega_{\phi 0} = 0.7$, the linear fit to equation (\ref{wpred})
for $z \la 1$ is roughly $w = w_0 - 1.5(1+w_0)(1-a)$, so that
$w_a \approx -1.5(1+w_0)$.

Now we can compare the generic behavior predicted by equation
(\ref{wpred}) with the actual scalar field evolution.  In Fig. 4,
we show this predicted behavior, along with $w(a)$ for
the potentials $V = \phi^2$, $V = \phi^{-2}$, and
$V = \exp(-\lambda \phi)$.  The value of $\phi_0$
for the power law potentials is chosen to give
$\Omega_{\phi 0} = 0.7$ and $w_0 = -0.9$.  For the exponential
potential, $\Omega_{\phi 0}=0.7$ and $w_0=-0.9$ are fixed by the value
of $\lambda$.
\begin{figure}[t]
\centerline{\epsfxsize=3.7truein\epsfbox{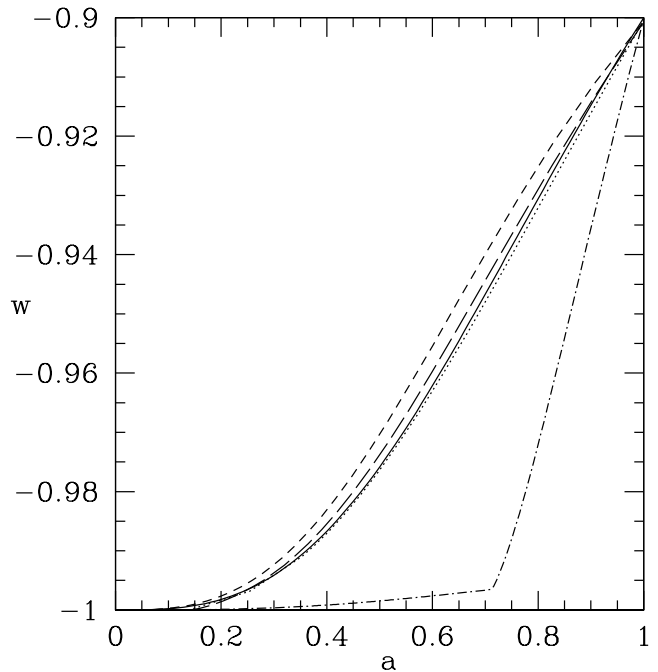}}
\caption{The evolution of $w$ as a function of the scale factor,
$a$, normalized to $a=1$ at the present, with $\Omega_{\phi 0} = 0.7$
and $w_0 = -0.9$.  Solid
curve is our analytic result for the behavior
of thawing models with a nearly flat potential and $w$ near $-1$.  Other curves
give the true evolution for the potentials
$V(\phi) = \phi^2$ (dotted), $V(\phi) = \phi^{-2}$ (short dash),
and $V(\phi) = \exp(-\lambda\phi)$ (long dash).  Dot-dash curve
is a model with $V(\phi) = \exp(-\lambda\phi)$ in which
$\lambda$ changes discontinuously.}
\end{figure}
The typical errors here are $\delta w \la 0.005$,
showing strong agreement between the true evolution of $w(a)$ and
our analytic expression for $w(a)$.  The error decreases as $w_0$ decreases, so
Fig. 4 gives an upper limit on the error in our approximation
for $w_0 < -0.9$.

In Fig. 5, we compare our limiting behavior for $w(a)$ in equation
(\ref{wpred}) to the SNIa observations.
The likelihoods were constructed using the 60 Essence supernovae, 57 SNLS (Supernova
Legacy Survey) and 45 nearby supernovae, and the new
data release of 30 SNe Ia detected by HST and classified as the Gold sample by
Riess et al. \cite{Riess1,Riess2}.  The combined dataset can be found in Ref. \cite{Davis}.
It is clear that current observations do not exclude the thawing quintessence
models we have considered here, although the observations are also obviously
consistent with a cosmological constant.  The maximum likelihood point
actually lies below $w_0=-1$, but we have not extended our graph
to $w_0 < -1$, as our derivation of equation (\ref{wpred}) assumes
a standard quintessence model with $w \ge -1$ at all times.
\begin{figure}[t]
\centerline{\epsfxsize=3.7truein\epsfbox{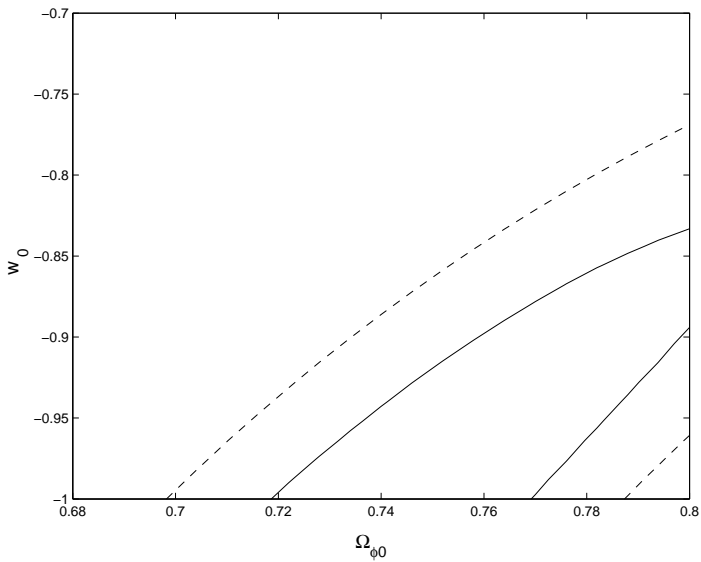}}
\caption{The $1-\sigma$ (solid) and $2-\sigma$ (dashed)
contours in the plane defined
by the present-day values of $\Omega_\phi$ and $w$ for
the quintessence field (denoted $\Omega_{\phi 0}$
and $w_0$ respectively), for the thawing behavior given
by equation (\ref{wpred}).}
\end{figure}

We have shown that the slow-roll conditions, equations (\ref{slow1})
and (\ref{slow2}), are sufficient to allow a thawing model
to produce $w$ near $-1$ today, but are they also necessary conditions?
The answer is no, although violating these bounds with a thawing model requires rather
unusual potentials.  Following Ref. \cite{Chong}, we have produced
a potential by joining the functions $V(\phi) = \exp(-\lambda_1 \phi)$
and $V(\phi) = \exp(-\lambda_2 \phi)$ at $\phi=0$.  The potential
is then continuous, but $\lambda$ varies discontinously, so
equation (\ref{slow2}) is violated.  In Fig. 4, we show $w(a)$
for this thawing potential (dot-dash curve), where we have
taken $\lambda_1 = 0.2$, $\lambda_2 = 1.04$, and we have
chosen $\phi_0$ to give $w_0 = -0.9$ and $\Omega_{\phi 0} = 0.7$.
The evolution of $\phi$ is clearly always in the thawing regime ($dw/da > 0$), and
it gives $w$ near $-1$, but it does not produce a functional form
for $w(a)$ resembling that of a nearly
flat potential.  Of course, this
form for the potential is rather pathological (note that it
also violates the upper bound on $w^\prime$ as a function of $w$ for thawing
models postulated
in Ref. \cite{CL}).

One of the best-motivated thawing quintessence models is the
Pseudo-Nambu Goldstone Boson (PNGB) model \cite{Frieman}.  (For a recent
discussion, see Ref. \cite{Albrecht} and references therein).
This model is characterized by the potential
\begin{equation}
V(\phi) = M^4 [\cos(\phi/f)+1],
\end{equation}
and $\phi_0$ can be taken to lie between $0$ and $\pi f$.  Then
the evolution of this model is a function of $M$, $f$, and $\phi_0$.
Using equations (\ref{slow1}) and (\ref{slow2}), we see that
the slow-roll conditions are satisfied for all $\phi_0$ if $f > 1$,
while they cannot be satisfied for any $\phi_0$ if $f < 1$.
The latter result follows from the trigonometic function in the
PNGB potential:  its second derivative is large whenever
the first derivative is small, and vice-versa.  None of these results
depend on the value of $M$.  Thus, our results are a good approximation
to the behavior of the PNGB model
for the case where $f >1$.

Now we consider some related approximation schemes.  Crittenden et al.
\cite{Crit} analyzed quintessence models with $w$ near $-1$ in
terms of the parameter $\kappa(\phi)$, defined through the equation
\begin{equation}
\kappa(\phi) = \frac{dV/d\phi}{V(1+\ddot\phi/3H\dot\phi)}.
\end{equation}
With $w$ near $-1$, they took the evolution for $\Omega_\phi$ to be given
by equation (\ref{Oma}), and they approximated the evolution of $\kappa$
as a linear function of $\phi$:
\begin{equation}
\kappa(\phi) = \kappa_0 + \kappa_1(\phi-\phi_0).
\end{equation}
Thus, the model of Ref. \cite{Crit} has two free parameters, $\kappa_0$ and $\kappa_1$,
which determine $w(a)$.  It is straightfoward to derive
the equivalent of our equations (\ref{mainresult}) and (\ref{wpred}).
For $w$ as a function of $\Omega_\phi$, we obtain:
\begin{equation}
\label{crit1}
1+w = \frac{2}{3} \kappa_0^2 \Omega_\phi
\left(\frac{1-\Omega_\phi}{1-\Omega_{\phi 0}}\right)^{4 \kappa_1/3}.
\end{equation}
Our corresponding result (equation \ref{mainresult}) is clearly
distinct from this result for all values of $\kappa_0$ and $\kappa_1$.
Note further that
equation (\ref{crit1}) implies $1+w_0 = (2/3)\kappa_0^2 \Omega_{\phi 0}$,
so we can express $w$ as a function of $w_0$, $\Omega_{\phi 0}$, and $\kappa_1$
alone (corresponding to equation \ref{wpred}); $\kappa_0$ drops
out of this expression:
\begin{equation}
\label{wCrit}
1+w = (1+w_0) a^3 \left[\Omega_{\phi 0} a^3 + (1-\Omega_{\phi 0})\right]^{-(4\kappa_1+3)/3}.
\end{equation}
Again, this result is distinct from equation (\ref{wpred}), although
the two expressions obviously can be made to converge to similar behavior by the
appropriate choice of $\kappa_1$, since $\kappa_1$ can be chosen to give
good agreement with the exact evolution \cite{Crit}.  The main difference
between this approach and ours is that our final result for $w(a)$
contains no free parameters; it is a function only of $w_0$ and $\Omega_{\phi 0}$,
while the expression for $w(a)$ in the form of equation (\ref{wCrit})
contains the fitting parameter $\kappa_1$.

Neupane and Scherer \cite{Neupane} considered the consequences of
fixing $x$ (as defined in equation \ref{xevol}) to be a constant, $\alpha$.
With $x = \alpha$, their relation corresponding to our equation
(\ref{mainresult}) is
\begin{equation}
\label{wNeup}
1+w = \frac{\alpha^2}{3 \Omega_\phi}.
\end{equation}
While it is possible in such models to produce $w$ close to $-1$ at
the present,
equation (\ref{wNeup}) shows that these $w \approx -1$ models always act as freezing
models, since increasing $\Omega_\phi$ corresponds to decreasing $1+w$.

One might argue that the correct ``generic" model for a nearly
flat potential should be a linear potential, i.e., constant $dV/d\phi$, rather
than constant $(1/V)(dV/d\phi)$.  Linear potentials
have been investigated previously by a number of authors
\cite{Linde,liddle,perivolaropoulos,doomsday,Sahlen1,Sahlen2}.
Exact solutions for the linear potential have been derived
for the case where the universe is scalar field dominated
\cite{Linde} or when it is matter-dominated \cite{liddle}, but
not for the intermediate case.  However, it is possible to use
the techniques discussed here to derive an approximate solution.
If we take
\begin{equation}
V = V_0 - \alpha \phi,
\end{equation}
then equation (\ref{motionq}) has the solution
\begin{equation}
\dot \phi = \alpha \int_{t = t_i}^{t_f} \left[\frac{a(t)}{a(t_f)}\right]^3
dt.
\end{equation}
The integrand can be reexpressed in terms of $a$ and $H(a)$ to
give
\begin{equation}
\label{linearphi}
\dot \phi = \alpha \int_{a=a_i}^{a_f} \frac{1}{H(a)}\left(\frac{a}{a_f}\right)^3
\frac{da}{a}.
\end{equation}
This solution is as yet exact.  Now we make essentially
the same approximations that we used
earlier for $1+w \ll 1$.  We take
$H(a)$ to be given by equation (\ref{H}), but in determining
the total value of $\rho$, we approximate $\rho_\phi$ as a constant,
and we take $1+w$ to be given by $1+w \approx \dot \phi^2/V(\phi_0)$.
With these approximations,
equation (\ref{linearphi}) can be used to derive $w(\Omega_\phi)$.
The resulting expression for $w(\Omega_\phi)$ is identical
to equation (\ref{mainresult}).  Furthermore,
numerical integration for the
linear potential gives results in good agreement
with our $(w_0, w_a)$ fit discussed above (see Fig. 4
of Ref. \cite{doomsday}). This supports the conclusion that our
results (equations \ref{mainresult} and \ref{wpred}) represent
a generic asymptotic behavior.  Of course, we cannot rule out
the possibility of a more exact solution for the linear potential than
the one we have outlined here.

\section{The case against thawing quintessence}

Now consider the arguments against the models considered here.  To
avoid unknown quantum gravity effects,
it is desirable for the energy scale of the scalar
field to be below the Planck mass (unity in our units).  Requiring
$\phi < 1$ does not constrain the models presented here, as the potential
can be modified to shift the value of $\phi$ to any desired value.
In Ref. \cite{CL}, it was suggested that a possible constraint
is $|V/V^\prime| < 1$.  Obviously, if this constraint is
enforced, then all of the models considered here are ruled out, since
we have only considered models with $V^\prime/V < 1$ (equation \ref{slow1}).
The implications
of this proposed constraint are explored further in Ref. \cite{Linder}.
In Ref. \cite{Huang}, it was argued that the correct constraint
is actually $\Delta \phi < 1$, where $\Delta \phi$ is the change
in the value of $\phi$ from its initial value to the present.
Our models do satisfy this second constraint.

Linder \cite{Linder} has noted that thawing models with
$\lambda_0 \ll 1$ occupy only a very small fraction of the phase
plane defined by $w$ and $w^\prime$.  This is certainly true; if one
requires $w$ to be very close to $-1$ at present, and
assigns equal {\it a priori} weight to all phase trajectories in
the $w - w^\prime$ plane, then models with $\lambda_0 \ll 1$
are very unlikely.

Huterer and Peiris \cite{HP} performed Monte Carlo simulations
of quintessence models, sampling low-order polynomial potentials.
(See also the related work in Refs. \cite{Sahlen1,Sahlen2}).
They found essentially no acceptable thawing models were generated
using this procedure.  This is not surprising, since their procedure
samples a uniform distribution in the initial values of
both $[(1/V)(dV/d\phi)]^2$
and $(1/V)(d^2 V/d\phi^2)$, while our models require the initial
values of both of
these quantities to be much smaller than unity.

We do not dispute the conclusions in either Refs. \cite{Linder}
or \cite{HP}; they simply represent a different approach
to determining the most plausible models.  The models presented here
require a very flat potential.  One can argue
that the special nature of the potential makes such models
unlikely; however, we believe that
an observed value of $w$ near $-1$ argues in favor
of choosing such special potentials, while
the fact that all such models
converge to a similar evolution makes these models more interesting.
These thawing models do require a fine-tuning
of $\phi_0$; it must be chosen so that $V(\phi_0)$
is approximately equal to the dark energy density today.

The most serious problem with the models considered here is that
there is currently no compelling observational evidence to favor them
over a cosmological constant, as Fig. 5 shows.  On the other hand,
current observations do not rule out these thawing models.

\section{Conclusions}

Thawing models with potentials
that satisfy the slow-roll conditions provide a natural
way to produce $w$ near $-1$, and they
all converge to a single, universal behavior.  Such
models are, in some ways, the opposite of the tracker
models proposed in Ref. \cite{zlatev}.  The tracker
models are insensitive to the initial conditions, but
they depend sensitively on the shape of the potential
over the entire range of evolution of $\phi$.  The models
discussed here, in contrast, depend {\it only} on the initial
conditions, i.e., the value of $V$ and its derivatives
at $\phi_0$, but are insensitive to the shape of the rest
of the potential.  This situation arises because the field
never rolls very far along the potential, and so never has
a chance to ``see" the rest of the potential.

These models provide a very well-defined form for $w(a)$ that
depends only on the present-day values of $w$ and $\Omega_\phi$.
While we have provided a variety of arguments both for and against
such models, it is obvious that these issues will ultimately be
settled by observational data, rather than by the speculations
of theorists like us.

\acknowledgments

R.J.S. was supported in part by the Department of Energy (DE-FG05-85ER40226).
A.A.S. thanks the Universitat Aut\'{o}noma de Barcelona, where
part of this work was completed under grant UAB-CIRIT, VIS-2007.
We thank R. Crittenden, I. Neupane, F. Piazza, M. Sahl\'{e}n,
V. Sahni, and especially E. Linder for helpful comments on the manuscript.

\end{document}